\documentclass[final,5p,times,number,sort&compress]{elsarticle}
\usepackage[utf8x]{inputenc}  % Tildar (input)
\usepackage[T1]{fontenc} % Tildar (output)
\usepackage[english]{babel}
\usepackage{graphicx}  % needed for figures
\usepackage{mathbbol} % all ten digits with \mathbb
\usepackage{physics} % For physics mathematical notation
\usepackage{amssymb}
\usepackage{amsmath}
\usepackage[caption=false]{subfig} % the subfigure package is deprecated and the subcaption package is incompatible with elsarticle
\usepackage[colorlinks,hyperindex]{hyperref}
\hypersetup{colorlinks,citecolor=blue,linkcolor=blue,urlcolor=blue}
\addto\captionsenglish{} % so that appendixname works as expected in the elsarticle class
\newcommand{\A}{\operatorname{A}}
\newcommand{\B}{\operatorname{B}}
\newcommand{\C}{\operatorname{C}}
\newcommand{\D}{\mathfrak{D}}
\newcommand{\F}{\operatorname{F}}
\newcommand{\K}{\mathcal{K}}
\renewcommand{\L}{\mathcal{L}}
\renewcommand{\O}{\mathcal{O}}
\newcommand{\s}{\mathrm{s}}
\newcommand{\T}{\textsc{t}}
\DeclareMathOperator{\e}{\operatorname{e}}
\newcommand{\num}{\scriptstyle \operatorname{num}}
\newcommand{\SC}{\scriptscriptstyle \operatorname{SC}}
\newcommand{\WC}{\scriptscriptstyle \operatorname{WC}}
\begin{document}
\begin{frontmatter}
\title{Multiple-scale analysis of open quantum systems}
%\author[unal]{D. N. Bernal-García \fnref{fn1}\corref{cor1}}
\author[unal,udea]{D. N. Bernal-García \corref{cor1}}
\ead{dnbernalg@unal.edu.co}
\author[udea]{B. A. Rodríguez}
\author[unal]{H. Vinck-Posada}
\address[unal]{Grupo de Superconductividad y Nanotecnología, Departamento de Física, Universidad Nacional de Colombia, \\ Ciudad Universitaria, K. 45 No. 26-85, Bogotá D.C., Colombia}
\address[udea]{Grupo de Física Atómica y Molecular, Instituto de Física, Facultad de Ciencias Exactas y Naturales, Universidad de Antioquia, \\ Calle 70 No. 52-21, Medellín, Colombia}
%\fntext[fn1]{These authors contributed equally to this work.}
\cortext[cor1]{Corresponding author}
\date{\today}
\begin{abstract}
In this work, we present a multiple-scale perturbation technique suitable for the study of open quantum systems, which is easy to implement and in few iterative steps allows us to find excellent approximate solutions.
For any time-local quantum master equation, whether markovian or non-markovian, in Lindblad form or not, we give a general procedure to construct analytical approximations to the corresponding dynamical map and, consequently, to the temporal evolution of the density matrix.
As a simple illustrative example of the implementation of the method, we study an atom-cavity system described by a dissipative Jaynes-Cummings model.
Performing a multiple-scale analysis we obtain approximate analytical expressions for the strong and weak coupling regimes that allow us to identify characteristic time scales in the state of the physical system.
\end{abstract}

\end{frontmatter}
%
%------------------------------------------------------------------------------

%------------------------------------------------------------------------------
% Introduction
\section{Introduction}
\label{sec:intro}
The study of open quantum systems (OQS) has become mandatory when thinking about any technological application based on quantum mechanics \cite{Kavokin2007}. However, it is not limited to applications, but has been crucial in important contributions to fundamental physics \cite{Rivas2012}.
In fact, any area of physics interested in the quantum description of a system that interacts with its environment needs the theory of open quantum systems. Such is the case of quantum optics \cite{Walls2008}, quantum information \cite{Nielsen2000}, and quantum measurement \cite{Wiseman2010}, just to give some examples.
Therefore, finding appropriate mathematical methods to solve the dynamics of OQS is as relevant as the study of the physical systems themselves.
Traditionally, the dynamics and the steady state of an open quantum system are calculated by means of exact diagonalization of the Liouville superoperator (liouvillian) \cite{Quesada2011}, or by simulations with quantum trajectories \cite{Carmichael2008}. However, both approaches are very limited by the growth of the Hilbert space, and quickly become challenging numerical problems.
The challenge increases when the liouvillian is time-dependent, where in general it is not possible to make an exact analytical diagonalization to obtain the exponential solution to the quantum master equation.
In this case, it is common to use a Magnus series \cite{Magnus1954}, which is an analytical expression that allows expressing the exponential solution to a linear differential equation of linear operators as a series expansion.
Nevertheless, although it provides approximate analytical solutions, it is difficult to use due to the complicated structure of the mathematical expressions involved.
Therefore, it is necessary to look for other approaches to solve a wider spectrum of OQS that otherwise would be very difficult to study.
For this purpose, there are multiple more sophisticated analytical and numerical methods, including perturbative expansion of dissipative terms \cite{Yi2000, Fleming2011}, numerical renormalization group \cite{Anders2005,Bulla2008}, density matrix renormalization group \cite{Hartmann2009}, effective liouvillian superoperators \cite{Reiter2012, Kessler2012},
Feshbach projection technique \cite{Chruscinski2013}, variational matrix product operators \cite{Cui2015}, Keldysh Green functions \cite{Kamenev2011, Sieberer2016}, and Lindblad perturbation theory \cite{Li2014, Li2016}.
These methods are either focused on obtaining solutions to the steady state or are numerical techniques to solve the dynamics of OQS. In any case, they do not provide approximate analytical expressions that describe the dynamics of OQS in both the transient and steady states.
To this end, it is natural to consider a time-dependent perturbation theory; however, a usual perturbative expansion is often an insufficient approximation to the dynamics of classical and quantum systems, since it inevitably leads to secular terms in the approximate solutions, i.e., terms that grow as powers of the time variable, thus limiting the range of validity of the approximate analytical expressions.
Secular solutions appear in differential equations with constant coefficients, when the non-homogeneous term is itself a solution to the associated homogeneous differential equation, as it happens in the regular time-dependent perturbation theory \cite{Bender1999}.
In particular, in classical mechanics this type of asymptotic expansion provides secular solutions for systems known to evolve harmonically, and in quantum mechanics it gives secular terms that destroy the canonical commutation relations among approximate operators \cite{Janowicz2003}.
As an alternative to the regular time-dependent perturbation theory in OQS, we propose a perturbative approach based on the multiple-scale analysis (MSA) of time-local master equations (TLMEs). 
MSA methods have been around for a long time as techniques to avoid secular terms in the solution of ordinary differential equations. In fact, the first of these was the Poincaré-Von Zeipel method, which was developed at the end of the 19th century for the study of celestial dynamics \cite{Scherer1994}.
However, even though it provides approximate periodic solutions without secular terms, the asymptotic expansions do not converge and this limits their usability.
Nevertheless, this and other multiple-scale techniques were widely used, although the mathematical justification for its operation was not well understood until the 1960s, when multiple-scale modeling was explained using coordinate transformations and invariant manifolds \cite{Sandri1965, Ramnath1969}.
Since then, MSA has been used successfully in the solution of linear and weakly nonlinear differential equations describing the dynamics of classical systems \cite{Kevorkian1996, Bender1999, Nayfeh2004, Strogatz2014}; and more recently, solving the dynamics of  quantum systems, such as the quantum anharmonic oscillator \cite{Bender1996a, Bender1996b, Auberson2002}, a wide range of quantum optical systems \cite{Janowicz1997, Janowicz1998a, Janowicz1998b, Janowicz2003}, and time-dependent problems like the dynamical Casimir effect \cite{Dalvit1999, Crocce2001, Crocce2002, Dodonov2012}.
However, in the existing implementations of MSA techniques, the solutions obtained must be subject to additional constraints so that each term in the perturbative expansion is physically consistent \cite{Wang2007}. This limitation does not allow the construction of a general scheme, but for each physical system a different implementation must be made.
For example, in the MSA solution of the Heisenberg equation of motion for the quantum anharmonic oscillator \cite{Auberson2002}, the technique does not guarantee that the canonical commutation relations among the approximate creation and annihilation operators are satisfied, but in each iteration, the integration constants must be adjusted according to this requirement, which implies a step-by-step development that can hardly be generalized.
Nevertheless, as we will show, it is possible to construct a multiple-scale perturbation technique (MSPT) that is free of such restrictions, and still preserves the physical properties of the system at each step of the perturbative iteration.
Hence, the aim of this work is to extend MSA to the study of OQS described by TLMEs. Our treatment is not limited to a specific physical system; on the contrary, it is applicable to a wide range of perturbations and OQS.
Besides, it is not just a technique that avoids secular terms, but a tool to obtain approximate solutions of rapid convergence and which are stable with respect to initial conditions.
This paper is organized as follows: In Sec. \ref{sec:master_equations}, we describe time-local quantum master equations for the density matrix operator, which are the type of equations we focus on in this work. In Sec. \ref{sec:multiple_scale}, we describe the multiple-scale perturbation technique for open quantum systems. In Sec. \ref{sec:examples}, we give an illustrative example of the implementation of the method. Finally, in Sec. \ref{sec:conclusions}, we conclude.
%
%-------------------------------------------------------------------------------

%-------------------------------------------------------------------------------
% Master equations
\section{Time-local quantum master equations}
\label{sec:master_equations}
This section aims to provide a brief and clear definition of time-local quantum master equations, with the intention of making the document self-contained.
Readers familiar with the theory of open quantum systems can skip to the next section, where the perturbative technique is described in detail.
The study of OQS appears as a more realistic approach to the description of the quantum dynamics of systems that are interacting in a non-negligible way with an environment or reservoir.
Formally, system and reservoir constitute a universe that is itself a closed quantum system with Hilbert space $\mathcal{H}_T$, and evolving unitarily under the total hamiltonian $H_T(t)$.
Therefore, the evolution from an initial time $t_0$ to a posterior time $t$ of the total density matrix $\rho_T$ describing the physical state of this universe, will be described in general by the unitary operator $ U(t,t_0)$, such that ($\hbar=1$),
\begin{align}
    \rho_T(t) &=\, U(t,t_0)\, \rho_T(t_0)\, U^{\dagger}(t,t_0), \\
    U(t,t_0) &=\, \mathcal{T} \exp{ i\, \int_{t_0}^{t} H_T(t')\, \dd{t'} };
\end{align}
with $\mathcal{T}$ the Dyson time-ordering operator.
However, OQS would be very difficult to study if all degrees of freedom are considered. Thus, a partial trace is carried out over the reservoir, reducing the problem to an efficient set of parameters and operators describing the system dynamics,
\begin{equation}
    \rho(t)=\Tr_{R} \qty{ U(t,t_0)\, \rho_T(t_0)\, U^{\dagger}(t,t_0) }.
\end{equation}
If the reservoir is in a steady state with a given spectral decomposition $\rho_{R} = \sum_j\, P_j\, \ket{r_j} \bra{r_j}$, $\sum_j P_j = 1$, and the system and its environment are initially uncorrelated $\rho_T(t_0)=\rho(t_0)\otimes\rho_R$, we can explicitly trace over the reservoir and separate the evolution of the system,
\begin{align}
    \rho(t) = \sum\limits_{i,j} P_j\, \bra{r_i} U(t,t_0) \ket{r_j}\, \rho(t_0)&\, \bra{r_j} U^{\dagger}(t,t_0) \ket{r_i};
\end{align}
which can be written as,
\begin{align}
    \rho(t) = \sum\limits_{i,j} K_{i j}(t,t_0)\, \rho(t_0)&\,  K^{\dagger}_{i j}(t,t_0), \label{eq:kraus_1}
\end{align}
where it was defined $K_{i j}(t,t_0) = \sqrt{P_j} \bra{r_i} U(t,t_0) \ket{r_j}$.
Eq. \eqref{eq:kraus_1} is the Kraus decomposition of the system density matrix evolution \cite{Kraus1983}, with $K_{i j}(t,t_0)$ the Kraus operators satisfying the time-independent completeness relation \cite{Schaller2014},
\begin{align}\label{eq:kraus_2}
  \sum\limits_{i,j}\, & K^{\dagger}_{i j}(t,t_0) K_{i j}(t,t_0) =   \mathbb{1}.
\end{align}
It is clear, that the operator sum representation of the time-evolution in Eq. \eqref{eq:kraus_1} is not unitary in general, however, it must describe a valid physical time-evolution mapping a physical state $\rho(t_0)$ into a physical state $\rho(t)$ \cite{Bruss2008}; where a density matrix describing a state must be positive semidefinite and trace preserving \cite{Breuer2015}.
Therefore, the operator sum decomposition must represent a completely-positive trace-preserving (CPTP) dynamical map $\Phi(t,t_0)$ \cite{Kraus1971, Choi1975},
\begin{equation}\label{eq:cptp}
    \rho(t) = \Phi(t,t_0)\, \rho(t_0) = \sum\limits_{i,j} K_{i j}(t,t_0)\, \rho(t_0)\,  K^{\dagger}_{i j}(t,t_0).
\end{equation}
The complete positivity ensures that, for all $t>t_0$ in the solution domain, the density matrix describing the physical state of the system is positive semidefinite. This is a consequence of the representation theorem for quantum operations which states that, given a density matrix that evolves according to the operator sum representation in Eq. \eqref{eq:kraus_1} with the operators satisfying the completeness relation in Eq. \eqref{eq:kraus_2}, then the dynamical map $\Phi(t,t_0)$ describing its evolution is completely positive \cite{Gorini1976,Petruccione2007}. Meanwhile, the trace-preserving property,  $\Tr{\rho(t)} = \Tr{\rho(t_0)}=1$, is a consequence of the completeness relation for Kraus operators in Eq. \eqref{eq:kraus_2}.
In general, the time-evolution of the system density matrix shown in Eq. \eqref{eq:cptp} can be described by a TLME,
\begin{equation}\label{eq:tlme}
    \dot{\rho}(t) = \L(t)\, \rho(t),
\end{equation}
where the liouvillian $\L(t)$ is a linear non-hermitian superoperator and the infinitesimal generator of the dynamical map $\Phi(t,t_0)$ \cite{Andersson2007}. Therefore, it is satisfied,
\begin{align}
    \rho(t)= \Phi(t,t_0)\, \rho(t_0)= \Bqty{ \mathcal{T} \e^{ \int_{t_0}^{t} \L(t')\, \dd{t'}} }\, \rho(t_0).
\end{align}
It is important to note that the dynamics described by TLMEs can be distinguished between markovian and non-markovian, where markovianity is defined by means of the divisibility of the dynamical map \cite{Rivas2010,Rivas2014,Breuer2015}; where divisibility is defined as, $\Phi(t,t_0) =\, \Phi(t,s)\, \Phi(s,t_0)$ ($t \geq s \geq t_0$), for every $(t,s, t_0)$ in the solution domain.
In many applications, the Born-Markov approximation is justified and, consequently, to have master equations with time-independent liouvillians \cite{Walls2008},
\begin{equation}\label{eq:me_time_independent}
    \dot{\rho}(t) = \L\, \rho(t).
\end{equation}
Thus, $\rho(t_1) =\, \e^{\L (t_1-t_0)}\, \rho(t_0)$ and the evolution will be homogeneous in time given by the single parameter, $t=t_1-t_0$.
Hence, the solution to Eq. \eqref{eq:me_time_independent} is given by,
\begin{equation}
    \rho(t) = \Phi(t)\, \rho(0) = \e^{\L\, t}\, \rho(0),
\end{equation}
with $t_0=0$ and, therefore, the system dynamics is represented by a quantum dynamical semigroup, i.e., a single-parameter dynamical map $\Phi(t)$ satisfying the semigroup property $\Phi(t+s)=\Phi(t)\, \Phi(s)$ ($t,s \geq 0$) \cite{Alicki2007}.
The most general type of markovian and time-homogeneous master equation describing non-unitary dynamics of the density matrix $\rho(t)$, is the celebrated Gorini-Kossakowski-Lindblad-Sudarshan (GKLS) master equation, or simply, the master equation in Lindblad form \cite{Gorini1976,Lindblad1976},
\begin{align}
    &\dot{\rho}(t) =  \L_{\mathrm{GKLS}}\, \rho(t) = -i\, \bqty{H, \rho(t)} + \sum\limits_n \gamma_n\, \mathcal{D}\bqty{O_n}\, \rho(t), \label{eq:lindblad:1} \\
    \mathcal{D}&\bqty{ O_n }\, \rho(t) = O_n\, \rho(t)\, O_n^{\dagger} - \frac{1}{2}\left( O_n^{\dagger}\, O_n\, \rho(t) + \rho(t)\, O_n^{\dagger}\, O_n \right),
\end{align}
where $H$ is the system's hamiltonian, and $\gamma_n$ are the parameters associated with the different dissipative and decoherent processes described by the operators $O_n$. Besides, if $ \gamma_n > 0 $, then the dynamical map $ \Phi(t) = \e^{\L_{\mathrm{GKLS}} \, t} $ is completely-positive.
A detailed derivation of the Lindblad master equation in Eq. \eqref{eq:lindblad:1} can be found in Refs. \cite{Carmichael1999, Walls2008}.
On the other hand, when the Born-Markov approximation can not be made, we are left with a time-dependent generator $\L(t)$, which has been shown that in general can be brought into a canonical Lindblad form with time-dependent coefficients and operators \cite{Hall2014,Hush2015},
\begin{align}\label{eq:lindblad:2}
  &\dot{\rho} = \, -i\, \bqty{H(t), \rho} \nonumber \\
  &+ \sum\limits_n \gamma_n(t)\, \Bqty{O_n(t)\, \rho\, O_n^{\dagger}(t) - \frac{1}{2}\left( O_n^{\dagger}(t)\, O_n(t)\, \rho + \rho\, O_n^{\dagger}(t)\, O_n(t) \right)},
\end{align}
The rates $\gamma_n(t)$ can only be temporarily negative in order to preserve the complete positivity of the map, in addition,  $\gamma_n(t) \geq 0$ for all $t$ in the solution domain, is a necessary and sufficient condition for the divisibility of the quantum operation and, consequently, for the markovianity of the system it describes \cite{Breuer2015}.
The above description is based on the assumption of an initially uncorrelated state between system and reservoir; however, in a more general scenario, a standard approach to the dynamics of OQS leads to the Nakajima-Zwanzig nonlocal quantum master equation,
\begin{equation}\label{eq:nonlocal}
    \dot{\rho}(t) = \int_{t_0}^{t} \mathcal{K}(t-t')\, \rho(t') \dd{t'} ,
\end{equation}
where $\mathcal{K}(t)$ is a memory kernel that accounts for the memory effects of the system, i.e., how the rate of change of the state at a given time $t$ depends on the physical state at all previous times. 
The nonlocal master equation \eqref{eq:nonlocal} represents a challenge when looking for solutions, even approximate ones, nevertheless, through the time convolution-less projection operator technique it is possible to bring a nonlocal master equation to a time-local form, $\dot{\rho}(t) = \mathcal{L}(t)\, \rho(t)$; although, this approach often leads to significant problems, such as the lack of complete positivity of the corresponding dynamical map \cite{Petruccione2007}.
However, an alternative that leads to perfectly regular dynamics was proposed in Ref. \cite{Chruscinski2010}, where it was shown that any solution to the nonlocal Eq. \eqref{eq:nonlocal} always satisfies a TLME with a generator that depends explicitly on the initial time, $\dot{\rho}(t) = \mathcal{L}(t-t_0)\, \rho(t)$.
Note that as a consequence of the aforementioned, it is not possible in general to reduce the dynamical map $\Phi(t,t_0)$ to a single-parameter representation, i.e., to a quantum dynamical semigroup.
Therefore, here we focus on systems whose dynamical map exists and is CPTP, without further additional conditions.
To simplify the notation, we assume that the initial time is always zero ($t_0 = 0$) and then, $\Phi(t,t_0=0) = \Phi(t)$; although, in general, we do not work with single-parameter dynamical maps.
In summary, the description of the dynamics of OQS in terms of TLMEs and CPTP dynamical maps is based in principle on the assumption of an initially uncorrelated state between system and reservoir. Nevertheless, by using the appropriate techniques, more general OQS can also be described by TLMEs.
On the other hand, the markovianity of the system is defined by means of the divisibility of the dynamical map; thus, if the liouvillian $\L$ is time-independent, then, the dynamics is trivially markovian; on the other hand, if $\L$ is time-dependent, the markovianity depends on whether the map is divisible or not.
However, the technique presented in this work, is suitable to obtain approximate analytical solutions to TLMEs describing both markovian and non-markovian dynamics.
%
%-------------------------------------------------------------------------------

%-------------------------------------------------------------------------------
% Multiple-scale analysis
\section{Multiple-scale perturbation technique}
\label{sec:multiple_scale}
%
%------------------------------------------------------------------------------
\begin{figure*}[p]
    \centering
    \includegraphics[width=0.8\textwidth]{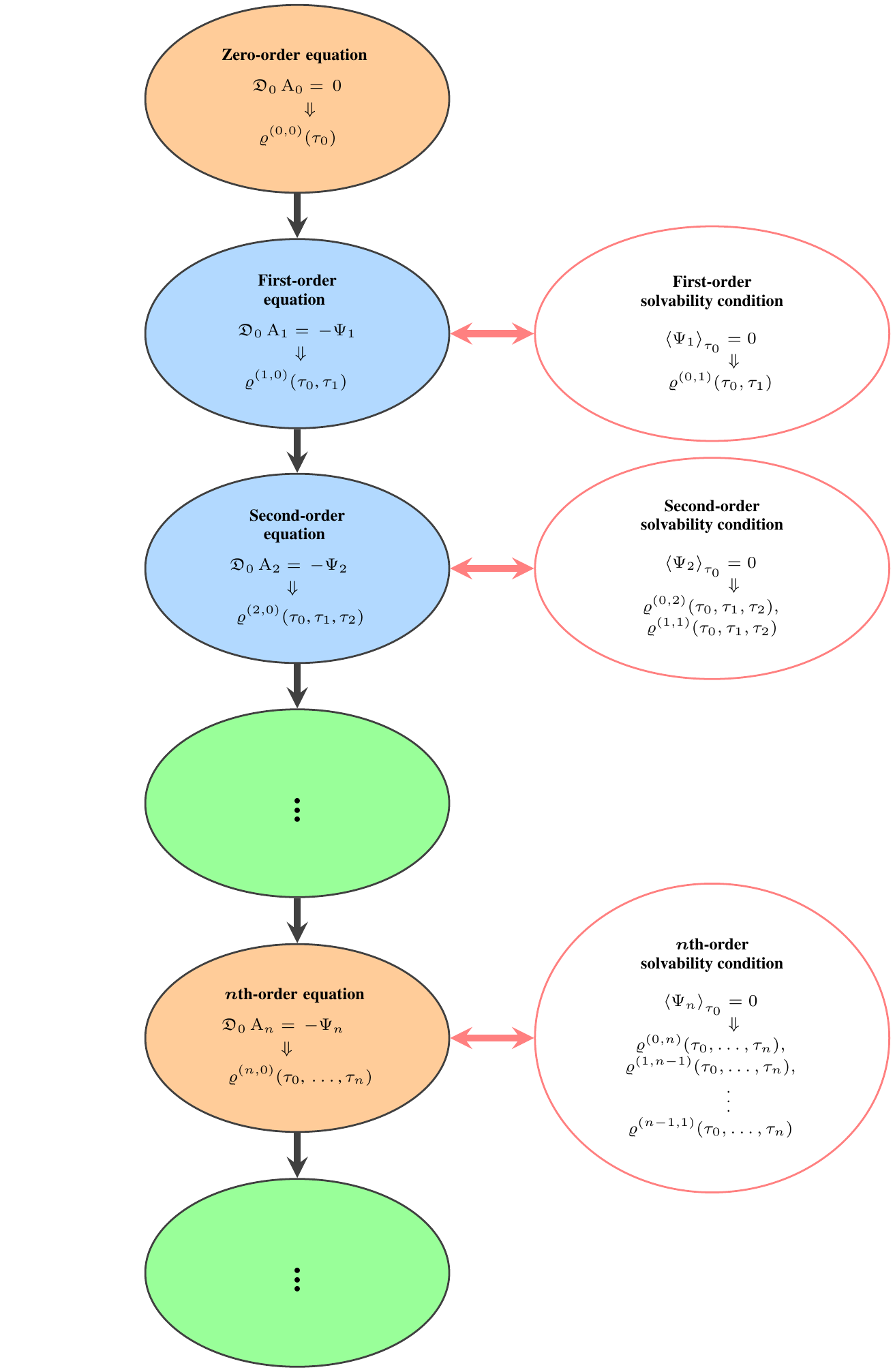}
    \vspace{-0.1cm}
    \caption{Flowchart schematizing the solution algorithm. We solve in each step first the solvability condition associated with the equation to solve. The solvability conditions will intercalate the equations for the different orders, and must be fulfilled if we require absence of secular terms. \label{fig:flowchart}}
\end{figure*}
%------------------------------------------------------------------------------
%

%
As pointed out above, the key goal of this research is to illustrate a perturbation technique based on the multiple-scale analysis of open quantum systems whose evolution is represented by time-local quantum master equations,
\begin{equation}
    \dot{\rho}(t) = \L(t)\, \rho(t).
\end{equation}
Such MSPT is essentially a derivative expansion method \cite{ Bender1999, Auberson2002}, which promotes the time variable $t$ to an infinite collection
$\T = \qty{\tau_0, \tau_1, \tau_2, \dots}$  of independent variables $\tau_n$, each one a function of the perturbative parameter $\alpha$ such that $\tau_n = \alpha^n\, t, \quad n = 0, 1, 2, \dots$, with $ 0<\alpha<1$. In consequence, the temporal derivative transforms into an infinite sum provided by the chain rule,
\begin{align}
  \dv{t} = \sum\limits_n \alpha^n \D_n, \qquad \D_n = \pdv{\tau_n}.
\end{align}
The permitted values of the perturbative parameter $\alpha$ guarantee the positivity of the time scales $\tau_n$ and their shortening with respect to the growth of the discrete variable $n$, i.e., whenever $n$ grows $\tau_n$ will describe increasingly rapid changes in the physical quantity that is being resolved.
In our implementation of MSA to TLMEs, we introduce an extended density matrix $\rho(\T) = \rho(\tau_0, \tau_1, \tau_2, \dots)$, considered an extension of the true density matrix $\rho(t)$, which is retrieved by restricting $\rho(\T)$: $ \rho(t) = \rho(\T)|_{\tau_n = \alpha^n t}$.
Expanding the density matrix asymptotically, we have,
\begin{align}
  \rho(\T) =& \sum\limits_n \alpha^n \varrho^{(n)}(\T).
\end{align}
It is worth mentioning that only $\varrho^{(0)}$ represents a valid density matrix.
In fact, it will be responsible of the unit-trace (all the other terms of the expansion are traceless), and the satisfaction of initial conditions,
$\varrho^{(0)}(0)=\rho(0)$ (while $\varrho^{(n)}(0)=0$ for $n>0$).
On the other hand, we can separate the liouvillian $\L(t)$ in two parts: A solvable superoperator $\L_0$ and a perturbation $\L_1(t)$, such that $\L(t) = \L_0+\alpha\,\L_1(t)$; besides, in the extended time representation we would have, $\L(t) \rightarrow \L(\tau_0)$; thus,
$	\L(\tau_0) = \L_0+\alpha\,\L_1(\tau_0) $. 
Putting the above expressions together in the TLME,
\begin{align}
  \dot{\rho}(\T) &= \L(\tau_0)\, \rho(\T), \nonumber \\
  \sum\limits_{m,n} \alpha^{m+n}\, \D_m\, \varrho^{(n)}(\T) &= \pqty{ \L_0 + \alpha \L_1(\tau_0)\, }\, \sum\limits_n  \alpha^n\, \varrho^{(n)}(\T);
\end{align}
and separating according to the powers of the perturbation parameter,
\begin{subequations}
\begin{align}
	\O(\alpha^0&): \D_0 \varrho^{(0)} - \L_0 \varrho^{(0)} = 0, \label{eq:homogeneous_eqn} \\
	\O(\alpha^1&):  \D_0 \varrho^{(1)} - \L_0 \varrho^{(1)} =  %\nonumber \\ & \hspace{1.6cm} 
	- \left( \D_1  \varrho^{(0)} -  \L_1(\tau_0) \varrho^{(0)} \right), \\
	\O(\alpha^2&): \D_0 \varrho^{(2)} - \L_0 \varrho^{(2)} = \nonumber \\ 
	& \hspace{0.7cm}  - \left( \D_1 \varrho^{(1)} - \L_1(\tau_0) \varrho^{(1)} \right)  - \D_2 \varrho^{(0)},  \\ 
	&  \hspace{3.1cm} \vdots \nonumber \\
	\O(\alpha^n&): \D_0 \varrho^{(n)} - \L_0 \varrho^{(n)} =  \nonumber \\ 
	& \hspace{-0.45cm} - \left( \D_1 \varrho^{(n-1)} - \L_1(\tau_0)  \varrho^{(n-1)} \right) - \sum\limits_{m=0}^{n-2} \D_{n-m} \varrho^{(m)}, \nonumber \\
	& \hspace{4.5cm} n \geq 2;
\end{align}
\label{eq:perturbative_eqs_1}
\end{subequations}
we can determine the time-evolution of the density matrix by means of the iterative solution of Eqs. \eqref{eq:perturbative_eqs_1}.
Hence, from the zero-order equation \eqref{eq:homogeneous_eqn}, performing a Dyson integration as shown in \ref{app:homogeneous_equation},
\begin{equation}\label{eq:zero_order_rho}
\varrho^{(0)}(\T) = \Bqty{ \e^{\L_0 \tau_0} \C_{0,1}(\T_1) }\, \varrho^{(0)}(0),
\end{equation}
where $\T_1$ follows the notation, $\T_k = \qty{\tau_k, \tau_{k+1}, \dots}$ $(k \geq 1)$, i.e., the new collection $\T_k$ includes all the time-scales from $\tau_k$.
Besides, $\C_{0,1}(\T_1)$ is a constant superoperator with respect to the $\tau_0$ variable, whose dependence with respect to each time variable in $\T_k$ will be determined order-by-order in the perturbation theory.
Eq. \eqref{eq:zero_order_rho} suggests that the description of the perturbative technique can also be performed by means of the dynamical map $\Phi(\T)$, then if,
\begin{align}
  \varrho^{(0)}(\T) = \Phi^{(0)}(\T)\, \rho(0) = \Bqty{ \e^{\L_0 \tau_0} \C_{0,1}(\T_1) }\, \rho(0),
\end{align}
we can make describe the MSPT regardless of the initial conditions $\rho(0)$.
However, a description in terms of the extended dynamical map $\Phi(\T)$ will require to represent it as an asymptotic expansion as well,
\begin{align}
  \Phi(\T) = \sum\limits_n \alpha^n \Phi^{(n)}(T),
\end{align}
where only $\Phi^{(0)}(\T)$ represents a valid dynamical map satisfying the initial condition $\Phi^{(0)}(0)=\mathbb{1}$, while  $\Phi^{(n)}(0)=\mathbb{0}$ for $n>0$.
The time-evolution of the system density matrix will be given by $\rho(\T) = \Phi(\T)\, \rho(0)$, and therefore,
\begin{align}
  \rho(\T) = \sum\limits_n \alpha^n \Phi^{(n)}(\T)\, \rho(0)&, \\
  \varrho^{(n)}(\T) =  \Phi^{(n)}(\T)\, \rho(0)&.
\end{align}
Besides, we can separate the unperturbed solution from the expansion, which is analogous to moving to an interaction picture,  $ \Phi^{(n)}(\T) = \e^{\L_0 \tau_0} \A_n(\T)$, and then,
\begin{equation}
  \varrho^{(n)}(\T) =  \Bqty{e^{\L_0 \tau_0}  \A_n(\T)}\, \rho(0).
\end{equation}
In this way, in terms of the  $\A_n = \A_n(\T)$, Eqs. \eqref{eq:perturbative_eqs_1} transform to,
\begin{subequations}\label{eq:perturbative_eqs_2}
    \begin{align}
	    \O(\alpha^0):  \D_0\A_0 =& \mathbb{0}, \label{eq:perturbative_eqs_2:1}  \\
	    \O(\alpha^1): \D_0\A_1 =& - \left( \D_1 \A_0 - \F_{1,0}(\tau_0) \A_0 \right), \label{eq:perturbative_eqs_2:2}   \\
	    \O(\alpha^2): \D_0\A_2 =& - \left( \D_1\A_1 - \F_{1,0}(\tau_0) \A_1 \right) %\nonumber \\ &\hspace{3.3cm} 
	    - \D_2\A_0, \label{eq:perturbative_eqs_2:3}    \\
	    \vdots  \hspace{0.1cm}  & \nonumber \\
	    \O(\alpha^n): \D_0\A_n =& - \left( \D_1 \A_{n-1} - \F_{1,0}(\tau_0) \A_{n-1} \right) \nonumber \\
	    & \hspace{1.9cm} - \sum\limits_{m=0}^{n-2} \D_{n-m}\A_m, \nonumber \\
        & \hspace{1.9cm}	\quad n \geq 2; \label{eq:perturbative_eqs_2:4}
    \end{align}
\end{subequations}
where, $\F_{1,0}(\tau_0) = \e^{-\L_0 \tau_0}\, \L_1(\tau_0)\,\, \e^{\L_0 \tau_0}$.
The solution procedure will be iterative order to order, satisfying solvability conditions before solving each order equation.
Solvability conditions avoid the appearance of secular solutions, making zero the terms that resonate with the homogeneous equation.
We construct this conditions by making time averages with respect to the $\tau_0$ variable and separating the terms that oscillate with zero frequency, which are the ones that resonate with the homogeneous solution.
Hence, the solution to Eq. \eqref{eq:perturbative_eqs_2:1}, which gives us the $\tau_0$ dependence of $\A_0$, will be substituted into equation Eq. \eqref{eq:perturbative_eqs_2:2}, and here the solvability condition must be fulfilled before solving the first-order partial differential equation.
Thus, the first-order equation \eqref{eq:perturbative_eqs_2:2} can be written as $ \D_0\A_1 = -\Psi_1 $, with
$ \Psi_1 = \left( \D_1 \A_0 - \F_{1,0}(\tau_0) \, \A_0 \right)$; and the solvability condition will be, $\expval{\Psi_1}_{\tau_0} = \mathbb{0}$, which gives us the $\tau_1$ dependence of $\A_0$. After this, the improved first-order equation must be solved.
Therefore, the solvability conditions up to $n$th-order in the asymptotic expansion are,
\begin{subequations}
\begin{align}
	&\O(\alpha^1): \expval{\Psi_1}_{\tau_0} = \expval{ \D_1 \A_0 - \F_{1,0}(\tau_0) \A_0 }_{\tau_0} = \mathbb{0}, \label{eq:solvability:1}   \\
	&\O(\alpha^2): \expval{\Psi_2}_{\tau_0} = \expval{ \D_1\A_1 - \F_{1,0}(\tau_0) \A_1 + \D_2\A_0 }_{\tau_0} = \mathbb{0},  \label{eq:solvability:2}     \\
	& \hspace{2.55cm} \vdots \nonumber \\
	&\O(\alpha^n): \expval{\Psi_n}_{\tau_0} = \nonumber \\
	&\hspace{0.4cm} \expval*{ \D_1 \A_{n-1} - \F_{1,0}(\tau_0) \A_{n-1} + \sum\limits_{m=0}^{n-2} \D_{n-m}\A_m}_{\tau_0} = \mathbb{0}, \nonumber \\
    & \hspace{4.65cm} n \geq 2, \label{eq:solvability:n}
\end{align}
\label{eq:solvability}
\end{subequations}
where the time-averaging is given by,
\begin{align}
   \expval{\Psi_n}_{\tau_0} = \frac{1}{T_n}  \int_{0}^{T_n} \Psi_n \dd{\tau_0},
\end{align}
being $T_n$ the period of the solvability condition, which can approach infinity if there is no well-defined period.
In terms of the density matrix, it is possible to recognize the structure of the asymptotic sum up to a given level of precision. Thus, being $\rho^{(n)}$ the density matrix truncated to the $n$th-order, and $\varrho^{(m,n)}$ the $n$th correction to the $m$th term of the asymptotic expansion. For each order in the perturbative expansion, we have,
\begin{description}
\item[Zero-order:]
	$ \rho^{(0)}(\tau_0) = \varrho^{(0,0)}(\tau_0). $
\item[First-order:]
    $ \rho^{(1)}(\tau_0, \tau_1) =  \varrho^{(0,1)}(\tau_0, \tau_1) + \alpha\, \varrho^{(1,0)}(\tau_0, \tau_1). $
\item[Second-order:] $\rho^{(2)}(\tau_0,\tau_1,\tau_2) = $
    
    $\varrho^{(0,2)}(\tau_0, \tau_1, \tau_2) + \alpha\, \varrho^{(1,1)}(\tau_0, \tau_1, \tau_2) + \alpha^2\,  \varrho^{(2,0)}(\tau_0, \tau_1, \tau_2). $
\item[$n$th-order:]
	$ \rho^{(n)}(\tau_0,\dots,\tau_n) = $
	
	\hspace{2.32cm} $\sum\limits_{m=0}^n \alpha^m\, \varrho^{(m,n-m)} (\tau_0,\dots,\tau_{n}), \quad n \geq 1. $
\end{description}
Therefore, the solution algorithm will go as follows (as schematized in Fig. \ref{fig:flowchart}):
From the zero-order equation, Eq. \eqref{eq:perturbative_eqs_2:1}, $\varrho^{(0,0)}(\tau_0)$ is calculated; then, the non-homogeneous term $\Psi_1$ in the first-order equation, Eq. \eqref{eq:perturbative_eqs_2:2}, defines the first order solvability condition, $\expval{\Psi_1}_{\tau_0}=0$, which improves the previous solution providing the functional dependence with respect to the $\tau_1$ variable, $\varrho^{(0,1)}(\tau_0, \tau_1)$.
Next, the first order equation is solved, giving $\varrho^{(1,0)}(\tau_0,\tau_1)$. The approximate density matrix until this point is, $\rho^{(1)}(\tau_0, \tau_1) = \varrho^{(0,1)}(\tau_0, \tau_1) + \alpha\, \varrho^{(1,0)}(\tau_0,\tau_1)$.
In the same way, we can solve each order, improving the previous solution through the solvability condition, and then adding a new term to the asymptotic expansion.
On the other hand, when we return to the usual time variable $t$, we will have,
\begin{description}
  \item[Zero-order:]
  $ \rho^{(0)}(t) = \varrho^{(0,0)}(t) $
  \item[First-order:]
  $ \rho^{(1)}(t) = \varrho^{(0,1)}(t) + \varrho^{(1,0)}(t) $
  \item[Second-order:]
  $ \rho^{(2)}(t) = \varrho^{(0,2)}(t) + \varrho^{(1,1)}(t) + \varrho^{(2,0)}(t) $
  \item[$n$th-order:]
  $ \rho^{(n)}(t) = \sum\limits_{m=0}^n  \varrho^{(m,n-m)}(t), \quad n \geq 1. $
\end{description}
Now, we show the general expressions obtained from the solution to Eqs. \eqref{eq:perturbative_eqs_2} and \eqref{eq:solvability} for any open quantum system, with an arbitrary perturbation, described by a TLME.
\paragraph{Zero-order}
From Eq. \eqref{eq:perturbative_eqs_2:1}, as mentioned above, we have,
\begin{equation}
  \A_0(\T) = \C_{0,1}(\T_1).
\end{equation}
If we truncate to this order, the satisfaction of the initial conditions force that, $\C_{0,1}^{(0)} = \mathbb{1}$. Then, the zero-order density matrix will be,
\begin{equation}
  \rho^{(0)}(\tau_0) =\, \varrho^{(0,0)}(\tau_0),
\end{equation}
\begin{align}
  \varrho^{(0,0)}(\tau_0) =\, & \e^{\K_0 \tau_0} \rho(0), \qquad \K_0 = \L_0.
\end{align}
\paragraph{First-order}
The solvability conditions will provide the integration constants to the next level of precision, i.e., the functional dependence with respect to the next time variable.
Then, from the solvability condition in Eq. \eqref{eq:solvability:1}, we determine the $\tau_1$ dependence of $\C_{0,1}(\T_1)$, namely, the zero-order solution is improved to a next level of precision.
Now, looking after the first solvability condition, we are going to separate the superoperator $\F_{1,0}(\tau_0)$ in a part that oscillates with zero frequency,  $\expval{\F_{1,0}}_{\tau_0}$, which corresponds to the temporal average of the signal in a period $T_{1,0}$, and another, $\widetilde{\F}_{1,0}(\tau_0)$, that contains all the terms that oscillate with nonzero frequencies,
\begin{align}
  \F_{1,0}(\tau_0) &= \e^{-\K_0 \tau_0}\, \L_1(\tau_0)\, \e^{\K_0 \tau_0} \label{eq:f10} \\
  \F_{1,0}(\tau_0) &=  \expval{\F_{1,0}}_{\tau_0} + \widetilde{\F}_{1,0}(\tau_0);
\end{align}
\begin{equation}
  \expval{\F_{1,0}}_{\tau_0} = \frac{1}{T_{1,0}}  \int_0^{T_{1,0}} \F_{1,0}(\tau_0)\, \dd{\tau_0}.
\end{equation}
\begin{equation}
  \C_{0,1}(\T_1) =\e^{\K_1 \tau_1} \C_{0,2}(\T_2), \qquad \K_1 = \expval{\F_{1,0}}_{\tau_0}.
\end{equation}
Furthermore, $\A_1(\T)$ is calculated from Eq. \eqref{eq:perturbative_eqs_2:2}, and we have until first-order,
\begin{align}
  \A_0(\T) &=\, \C_{0,1}(\T_1)\, =\, \e^{\K_1 \tau_1}\, \C_{0,2}(\T_2), \\
  \A_1(\T) &=\,  \B_{1,0}(\tau_0)\, \C_{0,1}(\T_1) + \C_{1,1}(\T_1)\, \nonumber \\
  &=\, \B_{1,0}(\tau_0)\, \e^{\K_1 \tau_1}\, \C_{0,2}(\T_2) + \C_{1,1}(\T_1);
\end{align}
\begin{equation}
  \B_{1,0}(\tau_0) = \int_{0}^{\tau_0} \widetilde{\F}_{1,0}(\tau'_0) \dd{\tau'_0}.
\end{equation}
If we truncate to this order, the determination of the integration constants will be subject to the initial conditions, i.e., $\Phi(0)=\mathbb{1}$. Therefore, since $\B_{1,0}(0)=\mathbb{0}$ (in general, $\B_{m,n}(0)=\mathbb{0}$), $\C_{0,2}^{(1)}=\mathbb{1}$ and  $\C_{1,1}^{(1)}=\mathbb{0}$. Then, the first-order density matrix will be,
\begin{equation}
  \rho^{(1)}(\tau_0, \tau_1) = \varrho^{(0,1)}(\tau_0, \tau_1) + \varrho^{(1,0)}(\tau_0, \tau_1);
\end{equation}
\begin{align}
  \varrho^{(0,1)}(\tau_0,\tau_1) =&\, \e^{\K_0 \tau_0}\, \e^{\K_1 \tau_1}\, \rho(0), \\
  \varrho^{(1,0)}(\tau_0,\tau_1) =&\, \e^{\K_0 \tau_0}\, \B_{1,0}(\tau_0)\, \e^{\K_1 \tau_1}\, \rho(0).
\end{align}
\paragraph{Second-order}
From the second-order solvability condition in Eq. \eqref{eq:solvability:2}, the $\tau_2$ dependence of $\C_{0,2}(\T_2)$ and the $\tau_1$ dependence of $\C_{1,1}(\T_1)$ are provided. Then, defining the new quantity $\F_{2,0}(\tau_0)$, we have,
\begin{align}
  \F_{2,0}(\tau_0) =& \F_{1,0}(\tau_0) \B_{1,0}(\tau_0) - \B_{1,0}(\tau_0) \expval{\F_{1,0}}_{\tau_0} \\
  \F_{2,0}(\tau_0) =&  \expval{\F_{2,0}}_{\tau_0} + \widetilde{\F}_{2,0}(\tau_0);
\end{align}
\begin{equation}
  \expval{\F_{2,0}}_{\tau_0} = \frac{1}{T_{2,0}}  \int_0^{T_{2,0}} \F_{2,0}(\tau_0)\, \dd{\tau_0}.
\end{equation}
\begin{align}
  \F_{1,1}(\tau_1) =& \e^{- \K_1 \tau_1}\, \expval{\F_{2,0}}_{\tau_0}\, \e^{\K_1 \tau_1} \\
  \F_{1,1}(\tau_1) =& \expval{\F_{1,1}}_{\tau_1} + \widetilde{\F}_{1,1}(\tau_1);
\end{align}
\begin{equation}
  \expval{\F_{1,1}}_{\tau_1} = \frac{1}{T_{1,1}}  \int_0^{T_{1,1}} \F_{1,1}(\tau_1)\, \dd{\tau_1}.
\end{equation}
\begin{align}
  \C_{0,2}(\T_2) =& \e^{\K_2 \tau_2} \C_{0,3}(\T_3), \qquad  \K_2 = \expval{\F_{1,1}}_{\tau_1}, \\
  \C_{1,1}(\T_1) =& \e^{\K_1 \tau_1} \Bqty{ \B_{1,1}(\T_1) \C_{0,2}(\T_2) + \C_{1,2}(\T_2) }, \nonumber \\
  =&  \e^{\K_1 \tau_1} \Bqty{ \B_{1,1}(\T_1) \e^{\K_2 \tau_2} \C_{0,3}(\T_3) + \C_{1,2}(\T_2) };
\end{align}
\begin{equation}
  \B_{1,1}(\tau_1) = \int_{0}^{\tau_1} \widetilde{\F}_{1,1}(\tau'_1) \dd{\tau'_1}.
\end{equation}
In addition, the structure of the $\A_2(\T)$ superoperator is calculated from Eq. \eqref{eq:perturbative_eqs_2:3}, and then, until second-order,
\begin{align}
  \A_0(\T) =& \e^{\K_1 \tau_1} \e^{\K_2 \tau_2} \C_{0,3}(\T_3), \\
  \A_1(\T) =& \B_{1,0}(\tau_0) \Bqty{ \e^{\K_1 \tau_1} \e^{\K_2 \tau_2} \C_{0,3}(\T_3) } \nonumber \\
  &+ e^{\K_1 \tau_1} \Bqty{ \B_{1,1}(\T_1) \e^{\K_2 \tau_2} \C_{0,3}(\T_3) + \C_{1,2}(\T_2) }, \\
  \A_2(\T) =& \B_{2,0}(\tau_0) \C_{0,1}(\T_1)  + \B_{1,0}(\tau_0) \C_{1,1}(\T_1)   \nonumber \\ & \hspace{4.5cm} + \C_{2,1}(\T_1), \nonumber \\
  \A_2(\T) =& \B_{2,0}(\tau_0)  \e^{\K_1 \tau_1} \e^{\K_2 \tau_2} \C_{0,3}(\T_3) \nonumber \\
  & \hspace{-1cm} + \B_{1,0}(\tau_0) \e^{\K_1 \tau_1} \left\{ \B_{1,1}(\T_1) \e^{\K_2 \tau_2} \C_{0,3}(\T_3) \right. \left.  + \C_{1,2}(\T_2) \right\} \nonumber \\
  & \hspace{4.5cm} +  \C_{2,1}(\T_1);
\end{align}
\begin{equation}
  \B_{2,0}(\tau_0) = \int_{0}^{\tau_0} \widetilde{\F}_{2,0}(\tau'_0) \dd{\tau'_0}.
\end{equation}
If we truncate to this order, $\C_{0,3}^{(2)}=\mathbb{1}$, $\C_{1,2}^{(2)}=\mathbb{0}$, and $\C_{2,1}^{(2)}=\mathbb{0}$. Then, the second-order density matrix will be,
\begin{align}
  \rho^{(2)}(\tau_0, \tau_1, \tau_2) =& \varrho^{(0,2)}(\tau_0, \tau_1, \tau_2) + \varrho^{(1,1)}(\tau_0, \tau_1, \tau_2) \nonumber \\
  &+ \varrho^{(2,0)}(\tau_0, \tau_1, \tau_2);
\end{align}
\begin{align}
  \varrho^{(0,2)}(\tau_0,\tau_1,\tau_2) =& \e^{\K_0 \tau_0} \e^{\K_1 \tau_1} \e^{\K_2 \tau_2} \rho(0), \\
  \varrho^{(1,1)}(\tau_0,\tau_1,\tau_2) =& \e^{\K_0 \tau_0} \left\{ \B_{1,0}(\tau_0)  \e^{\K_1 \tau_1} \e^{\K_2 \tau_2} \right. \nonumber \\
  & \hspace{1cm} \left. + \e^{\K_1 \tau_1} \B_{1,1}(\tau_1) \e^{\K_2 \tau_2} \right\} \rho(0), \\
  \varrho^{(2,0)}(\tau_0,\tau_1,\tau_2) =& \e^{\K_0 \tau_0} \left\{\B_{2,0}(\tau_0) \e^{\K_1 \tau_1} \e^{\K_2 \tau_2} \right. \nonumber \\
  &\left. + \B_{1,0}(\tau_0) \e^{\K_1 \tau_1} \B_{1,1}(\tau_1) \e^{\K_2 \tau_2} \right\} \rho(0).
\end{align}
\paragraph{$n$th-order}
Similar to the previous steps, in the $n$th iteration, the $n$th solvability condition, Eq. \eqref{eq:solvability:n},  allows us to determine the temporal dependence to a higher order of all the superoperators that appeared previously, and then the $n$th-order equation \eqref{eq:perturbative_eqs_2:4} will allow finding the structure of the $\A_n(\T)$ superoperator.
On the other hand, we can find more general expressions if we define,
\begin{align}
  \C_{m,0}(\T) =\, \e^{\K_0 \tau_0} \A_m(\T).
\end{align}
Consequently, we determine the general structure of $\A_m(\T)$,
\begin{align}
 \A_m(\T) &= \sum\limits_{i=0}^m\, \B_{(m-i),0}(\tau_0) \C_{i,1}(\T_1), \qquad \B_{0,0}(\tau_n) =\, \mathbb{1};
\end{align}
\begin{align}
  \C_{m,n}(\T_n) =\,  \e^{\K_n \tau_n} \sum\limits_{i=0}^m\, \B_{(m-i),n}(\tau_n) \C_{i,n+1}(\T_{n+1})&, \nonumber \\
  \B_{0,n}(\tau_n) =\, \mathbb{1}&.
\end{align}
Hence, the superoperators $\B_{m,n}(\tau_n)$ ($m>0, n \geq 0$) and  $\K_n$ ($n>0$) can be calculated up to the desired order from the following expression, which summarizes Eqs. \eqref{eq:perturbative_eqs_2} and \eqref{eq:solvability},
\begin{align}
    &\sum\limits_{i=0}^{n-1}\, \D_0 \B_{n-i,0}(\tau_0) \C_{i,1}(\T_1) = \nonumber \\
    & \hspace{0.7cm} \F_{1,0}(\tau_0) \sum\limits_{i=0}^{n-1} \B_{n-i-1,0}(\tau_0) \C_{i,1}(\T_1) \nonumber \\
    & \hspace{0.3cm} - \sum\limits_{m=0}^{n-1} \sum\limits_{i=0}^{m} \B_{n-i-1,0}(\tau_0) \D_{n-m} \C_{i,1}(\T_1), \quad n \geq 1.
\end{align}
It is worth mentioning, as is shown in \ref{app:commutation}, that all superoperators $\K_n$ ($n \geq 0$) commute between them, $\comm{\K_{n+1}\, }{ \K_n } = \mathbb{0} $.
This means, that the zero term in the asymptotic expansion of the dynamical map,
\begin{equation}
    \Phi^{(0)}(\T) = \e^{\K_0 \tau_0}\,  \e^{\K_1 \tau_1}\, \dots \ \e^{\K_n \tau_n}\, \dots
\end{equation}
is an exponential factorization of commuting superoperators.
On the other hand, if $\L_0$ and $\L_1$ commute, the MSA solution will correspond to the exact solution, $\Phi(t) = \e^{\L_0 t}  \e^{\alpha \L_1 t}$; which is reached after solving the first-order solvability condition in the MSPT.
In the following section, we show a simple example of implementation of the MSPT, this with the intention of clarifying the ideas presented.
%
%-------------------------------------------------------------------------------
% Example
\section{Illustrative example: The dissipative Jaynes-Cummings model}
\label{sec:examples}
As a simple example of the implementation of the MSPT, let us consider the Jaynes-Cummings (JC) model with dissipation, which describes the quantum dynamics of a two-level system (TLS) interacting with a quantized single-mode of the electromagnetic field in a dissipative cavity.
The system-reservoir interaction occurs by means of the leakage of photons through the cavity mirrors and the continuous and incoherent pumping of the TLS, a situation common in semiconductor cavity quantum electrodynamics \cite{Khitrova2006}.
In principle, this problem could be solved formally using the Glauber-Sudarshan P-representation of the corresponding TLME \cite{Gardiner2004}, though, this task might be challenging and not free of approximations.
However, by performing a MSA we are able to obtain simple approximate analytical expressions for the strong and weak coupling regimes that allow us to study the competition between pumping and losses on the state of the physical system.
The JC hamiltonian has two parts, the first one contains the information of the energy of the system,
\begin{align}
	H_0 =\, \omega_c\, a^{\dagger} a + \frac{\omega_a}{2}\,  \sigma_z,
\end{align}
where, $a$ and $a^{\dagger}$ are the annihilation and creation photonic operators, $\sigma_z$ is the third Pauli matrix, and $\omega_c$ and $\omega_a$ are respectively the energies of the photonic mode (cavity) and the TLS (atom).
The second part, has the information of the interaction between the involved subsystems, which is characterized by the radiation-matter interaction constant $g$. Therefore, near resonance  ($\omega_c \approx \omega_a$), if $g$ is much smaller than the natural frequencies of the system ($\omega_c,\, \omega_a$), then, the rotating wave approximation (RWA) is justified and the interaction hamiltonian is,
\begin{align}
	g\, V =\, g\, \pqty{ a^{\dagger}\, \sigma_- + a\, \sigma_+ }.
\end{align}
Thus, the complete JC hamiltonian will be,
\begin{align}
  H =\, H_0\, +\,&  g\, V, \nonumber \\
	H =\,\omega_c\, a^{\dagger} a + \frac{\omega_a}{2}\,  \sigma_z +\,& g\, \pqty{ a^{\dagger}\, \sigma_- + a\, \sigma_+ }.
\end{align}
On the other hand, in presence of leakage of photons through the cavity mirrors and continuous and incoherent pumping of the TLS, the master equation in Lindblad form is,
\begin{align}
	\dot{\rho}(t) = -i\, \bqty{H, \rho(t)} + \kappa\,  \mathcal{D}\bqty{a}\, \rho(t) + P\, \mathcal{D} \bqty{\sigma_+}\, \rho(t),
\end{align}
where $\kappa$ is the rate of leakage of photons out of the cavity and $P$ is the amplitude of the continuous pump, and explicitly ($\rho=\rho(t)$),
\begin{align}
	\dot{\rho} = - i \, \bqty{ H, \rho }
  & + \frac{\kappa}{2}\, \qty{ 2\, a\, \rho\, a^{\dagger} - a^{\dagger}\,	a\, \rho - \rho\, a^{\dagger}\, a } \nonumber \\
  & + \frac{P}{2}\, \qty{ 2\, \sigma_+\, \rho\, \sigma_- - \sigma_-\, \sigma_+\, \rho - \rho\, \sigma_-\, \sigma_+ }.
\end{align}
If there is low pumping in the system ($P \ll g, \kappa$), according to the relationship between $g$ and $\kappa$ there are two clearly differentiated operating regimes \cite{Laussy2009, Vera2009, Lodahl2015}. The first one, known as the strong coupling (SC) regime, is characterized by the fact that the interaction constant is bigger than the system dissipation rate ($g > \kappa$). In the second one, known as the weak coupling (WC) regime, the opposite occurs, i.e., the interaction constant is smaller than the dissipation rate of the system ($g < \kappa$). For the description of any of these regimes, we assume that the atom can be in its ground $\ket{g}$ or excited $\ket{e}$ state, and the cavity photonic field can have zero $\ket{0_c}$ or one photon $\ket{1_c}$.
Therefore, the atom-cavity system is described by the bare states, $\ket{0} = \ket{g,0_c}$, $\ket{1} = \ket{e,0_c}$, $\ket{2} = \ket{g,1_c}$. The system of equations that describes the quantum dynamics in resonance ($\omega_c = \omega_a$) is the following,
\begin{subequations}
\begin{align}
	\dot{\rho}_{0,0} =\, & - P\, \rho_{0,0} + \kappa\, \rho_{2,2},  \\
	\dot{\rho}_{1,1} =\, & -i g \pqty{ \rho_{2,1} - \rho_{1,2} } + P\, \rho_{0,0},  \\
  \dot{\rho}_{2,2} =\, & - i g \pqty{ \rho_{1,2} - \rho_{2,1} } - \kappa\, \rho_{2,2},  \\
  \dot{\rho}_{1,2} =\, & - i g \pqty{ \rho_{2,2} - \rho_{1,1} } - \frac{\kappa}{2}\, \rho_{1,2};
\end{align}
\label{eq:system_equations}
\end{subequations}
where, $\rho_{j,i} = \rho_{i,j}^*$, and  $\dot{\rho}_{0,0}=-(\dot{\rho}_{1,1}+\dot{\rho}_{2,2})$ is a consequence of the trace-preservation.
We will consider that initially the atom is in its excited state and there are no photons in the cavity field, i.e., $\rho_{1,1}(0)=1$ and all other entries of the density matrix will be zero when $t=0$.
%
%-------------------------------------------------------------------------------

%
\subsection{Strong coupling}
%
%-------------------------------------------------------------------------------
%
\begin{figure*}[!t]
    \hspace{1.25cm}
    \subfloat{\label{fig:example_jc_populations:a} \includegraphics[width=.4\textwidth]{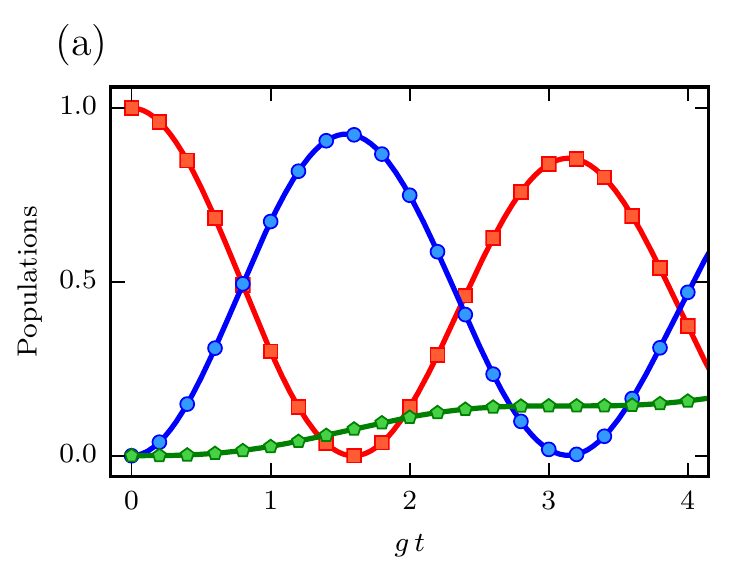} }
    %\hfill
    \hspace{1.25cm}
    \subfloat{\label{fig:example_jc_populations:b} \includegraphics[width=.4\textwidth]{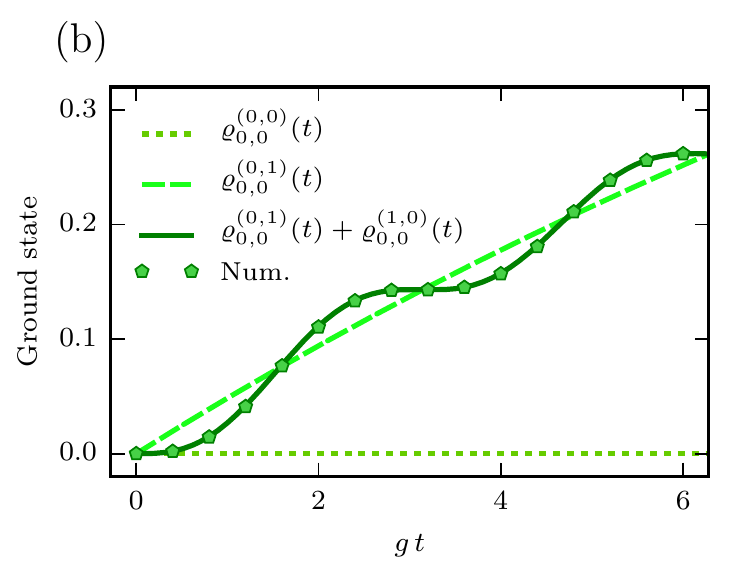}}
    \vspace{-0.4cm}
    \caption{Comparison between multiple-scale analysis and numerical integration solutions for the strong coupling regime. (a) Population of the states $\ket{0}$ (green with pentagon markers), $\ket{1}$ (red with square markers), $\ket{2}$ (blue with circle markers); the markers show the numerical solution, while the continuous lines represent the approximate solution to first-order in the multiple-scale perturbation technique. (b) Convergence of the approximations to the numerical solution for the population of the ground state $\ket{0}$; as before, the markers represent the numerical solution. For both plots, the parameters used were, $\omega_c = \omega_a = 1000\, g$, $\kappa = 0.1\, g$, $P = 0.01\, g$. \label{fig:example_jc_populations}}
\end{figure*}
\begin{figure*}[t]
    \hspace{1.25cm}
    \subfloat{\label{fig:example_jc_coherence:a} \includegraphics[width=.4\textwidth]{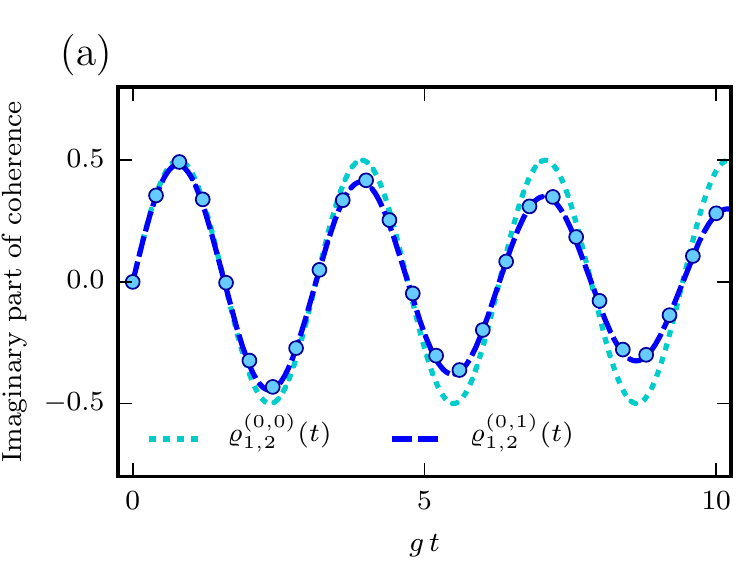} }
    \hspace{1.25cm}
    \subfloat{\label{fig:example_jc_coherence:b} \includegraphics[width=.4\textwidth]{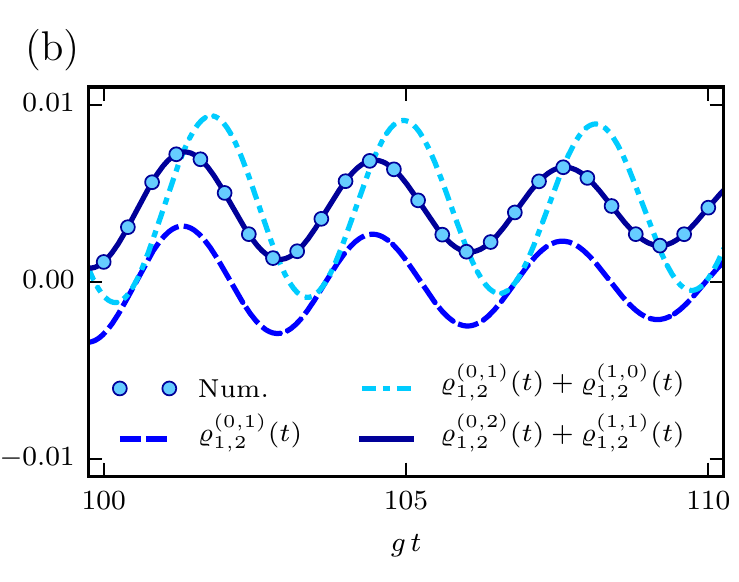}}
    \vspace{-0.4cm}
    \caption{Imaginary part of coherence in the strong coupling regime for, (a) short times, and, (b) long times. In both plots the circle markers represent the numerical solution to the system of ordinary differential equations. The parameters used were, $\omega_c = \omega_a = 1000\, g$, $\kappa = 0.1\, g$, $P = 0.01\, g$. \label{fig:example_jc_coherence}}
\end{figure*}
%
%-------------------------------------------------------------------------------
%

%
In the SC regime the perturbation will be associated with the Lindblad terms for incoherent pumping and leakage of photons through the cavity mirrors, then,
\begin{align}
	\L_0 =& -i\, \bqty{ H, \rho }, \\[0.2cm]
  \L_1 =\, & \kappa'\,  \mathcal{D}\bqty{a}\, \rho + P'\, \mathcal{D} \bqty{\sigma_+}\, \rho,
\end{align}
where, $\kappa = \alpha \kappa'$ and $P = \alpha P'$, being $\alpha$ the perturbation parameter.  Therefore, we will solve the master equation,
\begin{equation}
	\dot{\rho}(t) = \Bqty{ \L_0 + \alpha\, \L_1 }\, \rho(t).
\end{equation}
Afterwards, we will briefly describe the solution procedure up to second-order in the MSPT together with the approximate analytical expressions up to first-order in the MSPT. The shown analytical expressions were obtained using the general procedure developed in Sec. \ref{sec:multiple_scale}, and they were used to obtain the curves in Figs. \ref{fig:example_jc_populations} and \ref{fig:example_jc_coherence}.
\paragraph{Zero-order (SC)}
As mentioned above, the zero-order solution solves the unperturbed problem,

\begin{equation}
	\rho^{(0)}_{\SC}(t) = \varrho^{(0,0)}_{\SC}(t),
\end{equation}

\begin{align}
\varrho^{(0,0)}_{\SC}(t) &=
\begin{pmatrix}
0 & 0 & 0 \\
0 & \cos^2{(g t)} & \frac{i}{2} \sin{(2 g t)} \\
0 & -\frac{i}{2} \sin{(2 g t)} & \sin^2{(g t)} \\
\end{pmatrix}.
\end{align}
%

%
%-------------------------------------------------------------------------------
%
\begin{figure*}[!ht]
    \hspace{1.25cm}
    \subfloat{\label{fig:example_jc_wc:a} \includegraphics[width=.4\textwidth]{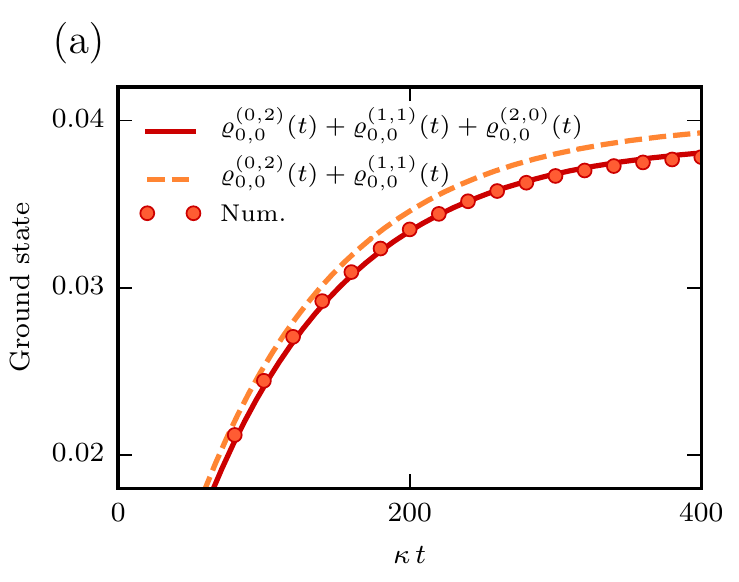} }
    \hspace{1.25cm}
    \subfloat{\label{fig:example_jc_wc:b} \includegraphics[width=.4\textwidth]{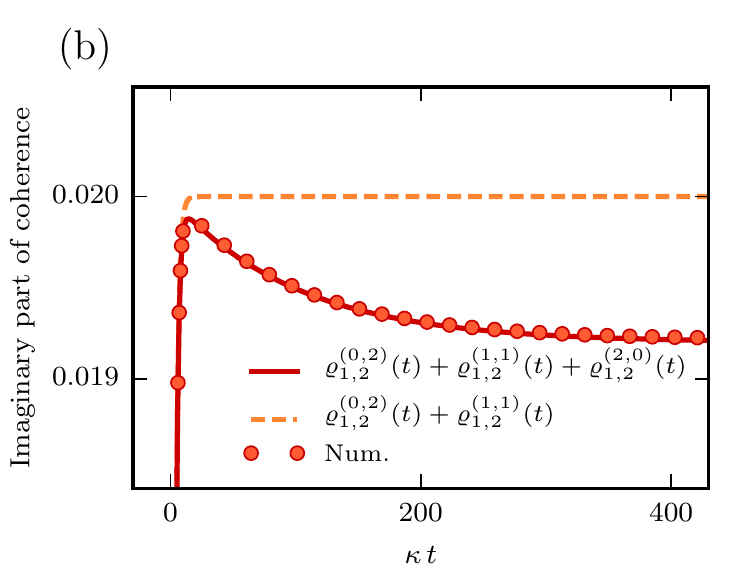}}
    \vspace{-0.4cm}
    \caption{Results of the multiple-scale perturbation technique in the weak coupling regime. (a) Population of the ground state. (b) Imaginary part of coherence. In both plots, the circle markers show the numerical solution. The parameters used were, $\omega_c = \omega_a = 1000\, \kappa$, $g = P = 0.01\, \kappa$. \label{fig:example_jc_wc}}
\end{figure*}
%
%-------------------------------------------------------------------------------
%

%
\paragraph{First-order (SC)}
To first-order in the MSPT, it is necessary to calculate $\F_{1,0}(\tau)$ and then its time-average $\expval{\F_{1,0}}_{\tau_0}$. For this purpose, the aforementioned superoperator was calculated and it was possible to recognize that its terms oscillate with a frequency $\Omega_{1,0}= g$; thus, the temporal average was calculated with respect to the period $T_{1,0}=2 \pi/g$. Using this result, we calculate  $\varrho^{(0,1)}_{\SC}(t)$ and $\varrho^{(1,0)}_{\SC}(t)$,
\begin{equation}
	\rho^{(1)}_{\SC}(t) = \varrho^{(0,1)}_{\SC}(t) + \varrho^{(1,0)}_{\SC}(t),
\end{equation}
\begin{subequations}
	\begin{flalign}
		& \varrho^{(0,1)}_{\SC_{0,0}}(t)= \frac{\kappa}{2P + \kappa} \pqty{ 1 - \e^{-(2P + \kappa)t/2} }, & \\
		& \varrho^{(0,1)}_{\SC_{1,1}}(t)= \frac{1}{2(2P + \kappa)} \left\{ 2P  + \kappa \e^{-(2P + \kappa)t/2} \right. & \nonumber \\
		&  \hspace{2.8cm} \left. + (2P + \kappa)\e^{-\kappa t/2} \cos{(2gt)} \right\}, & \\
		& \varrho^{(0,1)}_{\SC_{1,2}}(t)= \frac{i}{2} \e^{- \kappa t/2} \sin{(2 g t)}; &
	\end{flalign}
\end{subequations}
\begin{subequations}
	\begin{flalign}
		& \varrho^{(1,0)}_{\SC_{0,0}}(t)= -\frac{\kappa \sin (2 g t)}{4g}  \e^{-\kappa  t/2}, &  \\
		& \varrho^{(1,0)}_{\SC_{1,1}}(t)= \frac{\kappa \sin{(2gt)}}{{8 g (2P+\kappa)}} \left\{ (2P+\kappa) \pqty{ 1 + \e^{-P t} } \e^{-\kappa t/2} \right. & \nonumber \\
		& \hspace{3.3cm} \left. + 4P \pqty{1 - \e^{ -(2 P+\kappa) t/2} } \right\}, &  \\
		& \varrho^{(1,0)}_{\SC_{1,2}}(t)= \frac{i \kappa  \sin^2(g t)}{4 g (2P+\kappa)}  \left\{ (2P+\kappa) \e^{ -(2 P+\kappa) t/2} \right. & \nonumber  \\
		& \hspace{3.3cm} \left. + 4P \pqty{1 - \e^{ -(2 P+\kappa) t/2} } \right\}. &
	\end{flalign}
\end{subequations}
Up to first-order order in the perturbative expansion, the contribution to the frequency from $\kappa$ and $P$ appears simultaneously, providing the characteristic time scales, $2/\kappa$ and $2/(2 P+\kappa)$.
On the other hand, the approximate solution for the populations, up to this order is in very good agreement with the numerical solution, as is shown in Fig. \ref{fig:example_jc_populations}.
\paragraph{Second-order (SC)} To this order the time-averages were made with respect to the periods, $T_{2,0}=2 \pi/g$ and $T_{1,1}=2 \pi i/P'$.
Even though we do not show the approximate analytic expressions up to this order, it is important to note that a further improvement of the coherences was made through the second-order solvability condition, as plotted in Fig. \ref{fig:example_jc_coherence:b}.
In Fig. \ref{fig:example_jc_populations} and Fig. \ref{fig:example_jc_coherence}, we compare approximate and numerical solutions for the SC regime.
The parameters used in units of the coupling constant $g$ were, $\omega_c = \omega_a = 1000\, g$, $\kappa = 0.1\, g$, $P = 0.01\, g$.
The numerical solution was obtained by means of the numerical integration of the system of ordinary differential equations in Eqs. \eqref{eq:system_equations}, using the aforementioned parameters.
Fig. \ref{fig:example_jc_populations:a} presents the population in the transitory regime for the three states considered, it shows that the first-order solution in the MSPT is in perfect agreement with the numerical solution obtained. Besides, Fig. \ref{fig:example_jc_coherence:b} shows how the MSPT approaches the numerical solution for the population of the ground state.
On the other hand, Fig. \ref{fig:example_jc_coherence} shows the imaginary part of coherence between states $\ket{1}$ and $\ket{2}$. In particular, Fig. \ref{fig:example_jc_coherence:a} shows the perfect agreement of the first-order solvability condition solution with the numerical one for very short times.
In addition, Fig. \ref{fig:example_jc_coherence:b} shows that for long times this agreement is not fulfilled until the second-order solvability condition.
%
%-------------------------------------------------------------------------------
%

%
\subsection{Weak coupling}
Although it is natural to think that the perturbative term will always be associated with Lindblad dissipators, in some cases it is useful to include hamiltonian terms in the perturbation. Such is the case of the WC regime, where the perturbation will be associated with the hamiltonian interaction term and with the Lindblad term for continuous and incoherent pumping,
\begin{align}
	\L_0 =& -i\, \bqty{ H_0, \rho } + \kappa\,  \mathcal{D}\bqty{a}\, \rho, \\[0.2cm]
  \L_1 =&  -i\, g' \bqty{ V, \rho } + P'\, \mathcal{D} \bqty{\sigma_+}\, \rho,
\end{align}
where, $g = \alpha g'$ and $P = \alpha P'$ being $\alpha$ the perturbation parameter.
The plots of the approximate solutions for the WC regime are shown in Fig. \ref{fig:example_jc_wc}. The parameters used in terms of the dissipative rate $\kappa$ were, $\omega_c = \omega_a = 1000\, \kappa$, $g = P = 0.01\, \kappa$.
Next, we briefly describe the solution procedure together with the approximate analytical expressions up to second-order in the MSPT. As before, the shown analytical expressions were obtained using the general procedure developed in Sec. \ref{sec:multiple_scale}, and they were used to obtain the curves in Fig. \ref{fig:example_jc_wc}.
\paragraph{Zero-order (WC)} 
Unlike the previous case, the unperturbed solution does not show time-evolution of the system,
\begin{equation}
	\rho^{(0)}_{\WC}(t) = \varrho^{(0,0)}_{\WC}(t),
\end{equation}
\begin{align}
	\varrho^{(0,0)}_{\WC}(t)=
	\begin{pmatrix}
		0 & 0 & 0 \\
		0 & 1 & 0 \\
		0 & 0 & 0 \\
	\end{pmatrix}.
\end{align}
\paragraph{First-order (WC)} In this case the time-average of $\F_{1,0}(\tau_0)$  is performed with respect to an imaginary period $T_{1,0}=4 \pi i/\kappa$ in order to separate the zero-frequency terms.
To first-order in the MSPT,
\begin{equation}
	\rho^{(1)}_{\WC}(t) = \varrho^{(0,1)}_{\WC}(t) + \varrho^{(1,0)}_{\WC}(t),
\end{equation}
\begin{flalign}
	&\varrho^{(0,1)}_{\WC}(t)=
	\begin{pmatrix}
		0 & 0 & 0 \\
		0 & 1 & 0 \\
		0 & 0 & 0 \\
	\end{pmatrix},&
\end{flalign}

\begin{flalign}
	&\varrho^{(1,0)}_{\WC}(t)=
	\begin{pmatrix}
		0 & 0 & 0 \\
		0 & 0 & \frac{2 i g}{\kappa} \left(1 - \e^{-\kappa t/2}\right) \\
		0 & -\frac{2 i g}{\kappa} \left(1 - \e^{-\kappa t/2}\right) & 0 \\
	\end{pmatrix}.&
\end{flalign}
\paragraph{Second-order (WC)} 
To second-order the time-averages were calculated with respect to the periods, $T_{2,0} = 4 \pi i/\kappa$ and $T_{1,1}=2 \pi i/P'$. The second-order approximate solution is,
\begin{equation}
	\rho^{(2)}_{\WC}(t) = \varrho^{(0,2)}_{\WC}(t)+\varrho^{(1,1)}_{\WC}(t)+\varrho^{(2,0)}_{\WC}(t),
\end{equation}
\begin{flalign}
	&\varrho^{(0,2)}_{\WC}(t)=
	\begin{pmatrix}
		0 & 0 & 0 \\
		0 & 1 & 0 \\
		0 & 0 & 0 \\
	\end{pmatrix};&
\end{flalign}
\begin{subequations}
	\begin{flalign}
		& \varrho^{(1,1)}_{\WC_{0,0}}(t)=\,  \frac{4 g^2}{P \kappa} \pqty{ 1 - \e^{-P t} }, & \\
		& \varrho^{(1,1)}_{\WC_{1,1}}(t)=\,  - \frac{4 g^2}{P \kappa} \pqty{ 1 - \e^{-P t} }, & \\
		&\varrho^{(1,1)}_{\WC_{1,2}}(t)=\,  \frac{2 i g}{\kappa} \left(1-\e^{-\kappa t/2}\right); &
	\end{flalign}
\end{subequations}
\begin{subequations}
	\begin{flalign}
		& \varrho^{(2,0)}_{\WC_{0,0}}(t)=\,  -\frac{4 g^2}{\kappa^2} \pqty{ 3 - 4 \e^{-\kappa t/2} + \e^{-\kappa t} }, & \\
		& \varrho^{(2,0)}_{\WC_{1,1}}(t)=\, \frac{8 g^2}{\kappa^2} \pqty{ 1- \e^{-\kappa t/2} }, & \\
		&\varrho^{(2,0)}_{\WC_{1,2}}(t)=\, -\frac{8 i g^3}{P \kappa^2} \pqty{1-\e^{-P t} - \e^{-\kappa t/2} + \e^{-(2P+\kappa)t/2} }. &
	\end{flalign}
\end{subequations}
From comparison between Figs.
\ref{fig:example_jc_populations} and \ref{fig:example_jc_coherence}, and Fig. \ref{fig:example_jc_wc}; we can see that the results for the SC regime are in better agreement with the numerical calculations than those for the WC regime.
This is because in the latter case we have the perturbation of a purely damped system, which makes it difficult to introduce new time-scales that actually change the system dynamics.
Then, a better approximation for the WC regime would require going beyond in the perturbation expansion.
%
%-------------------------------------------------------------------------------

%
%-------------------------------------------------------------------------------
%
\begin{figure}[!t]
    \hspace{0.7cm}
    \includegraphics[width=.4\textwidth]{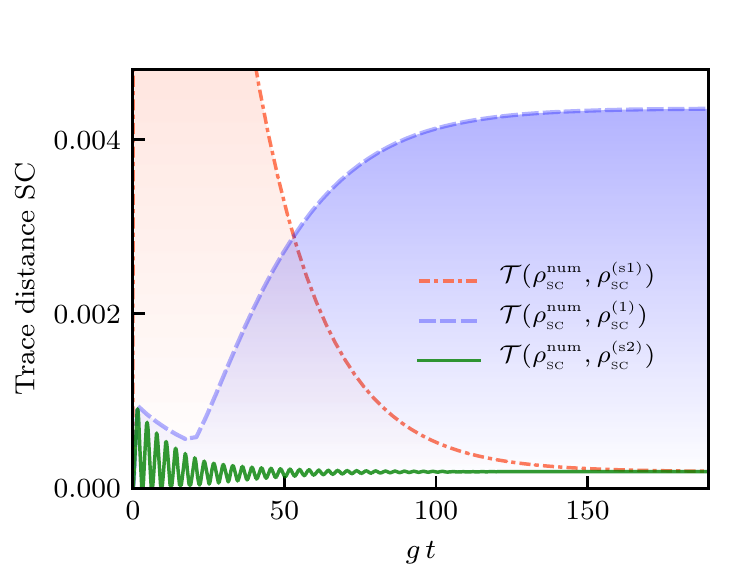}
    \vspace{-0.4cm}
    \caption{Trace distances between the numerical solution and different orders of the multiple-scale perturbation technique for the strong coupling regime.
    The three quantities plotted oscillate with the frequency of the Rabi oscillations; though, only the oscillations for $\mathcal{T}( \rho^{\num}_{\SC}, \rho^{\scriptscriptstyle (\s 2)}_{\SC}) $ are shown, whilst for $\mathcal{T}( \rho^{\num}_{\SC}, \rho^{\scriptscriptstyle (\s 1)}_{\SC}) $ and $\mathcal{T}( \rho^{\num}_{\SC}, \rho^{\scriptscriptstyle (1)}_{\SC}) $ the curves follow the local maxima.
    The parameters used here are the same as those used to obtain Figs. \ref{fig:example_jc_populations} and \ref{fig:example_jc_coherence}, $\omega_c = \omega_a = 1000\, g$, $\kappa = 0.1\, g$, $P = 0.01\, g$. 
    \label{fig:example_jc_td_sc}}
\end{figure}
\begin{figure}[t]
    \hspace{0.7cm}
    \includegraphics[width=.4\textwidth]{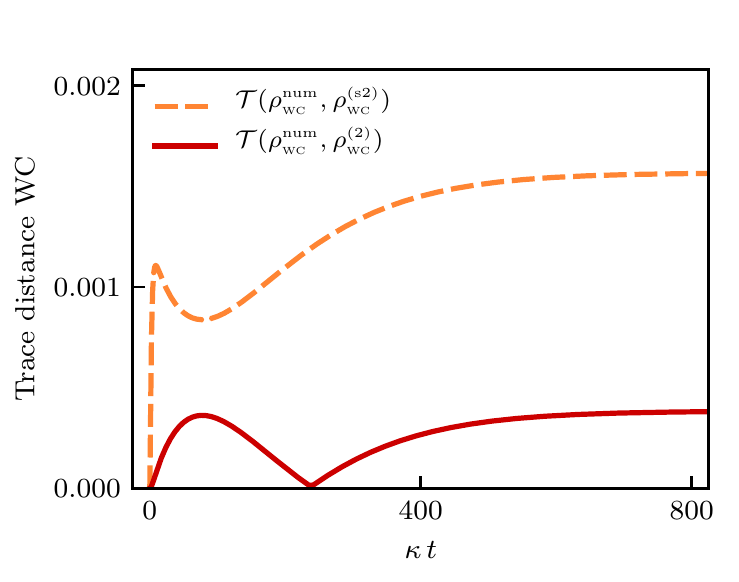}
    \vspace{-0.4cm}
    \caption{Trace distances between the numerical integration solution and different orders of the multiple-scale analysis in the weak coupling regime.
    The parameters used here are the same as those used to obtain Figs. \ref{fig:example_jc_wc}, $\omega_c = \omega_a = 1000\, \kappa$, $g = P = 0.01\, \kappa$. \label{fig:example_jc_td_wc}}
\end{figure}
%
%-------------------------------------------------------------------------------
%

%
\subsection{Error estimation}
In the quantum dynamics shown above, the truncation criterion was based on the qualitative comparison of the results obtained for each order of the perturbative expansion with the exact numerical dynamics. The asymptotic series were truncated when a good correspondence was found for short and long times according to the natural time scales of each problem.
Thus, the regimes of applicability of the method were not clear, since the errors introduced by the truncation process were not understood quantitatively.
Although, it is evident that ignoring higher order contributions will introduce significant errors in the approximation, especially, for long times, and perturbations $\alpha \L_1$ comparable with the unperturbed liouvillian of the system $\L_0$.
However, attempting to find an analytical estimation of errors essentially requires calculating the exact general evolution of the system density matrix in all regimes.
Nevertheless, using a combination of numerical and phenomenological approaches, here we provide a simple error estimation for the strong and weak coupling regimes, based on the results shown in Figs. \ref{fig:example_jc_populations}--\ref{fig:example_jc_wc}, thus quantifying the reliability of the found approximate analytical expressions.
To estimate the error and, therefore, arriving to more robust conclusions about accuracy and convergence to each order in the perturbative expansion, we use the trace distance between the quantum dynamics calculated numerically and the different orders in the perturbation theory for each regime.
The trace distance between two density matrices $\rho_1$ and $\rho_2$, is defined as,
\begin{equation}
    \mathcal{T} \pqty{\rho_1, \rho_2} = \frac{1}{2} \Tr{ \sqrt{\pqty{\rho_1 - \rho_2}^2 } } = \frac{1}{2} \sum_i \abs{ \lambda_i },
\end{equation}
where the $\lambda_i$ are the eigenvalues of the operator $(\rho_1 - \rho_2)$.
Figs. \ref{fig:example_jc_td_sc} and \ref{fig:example_jc_td_wc} show the trace distance for the strong and weak coupling regimes respectively, where we used the same sets of parameters as those in Figs. \ref{fig:example_jc_populations}--\ref{fig:example_jc_wc}.
In Fig. \ref{fig:example_jc_td_sc}, we study the time-evolution (extended towards long times) of the trace distance between the numerically calculated quantum dynamics and three different solutions in the MSPT.
Hence, we plotted,
\begin{flalign*}
    &\mathcal{T}\pqty{ \rho^{\num}_{\SC}(t), \rho^{(\s 1)}_{\SC}(t) },\,
    \mathcal{T}\pqty{ \rho^{\num}_{\SC}(t), \rho^{(1)}_{\SC}(t) },\,
    \mathcal{T}\pqty{ \rho^{\num}_{\SC}(t), \rho^{(\s 2)}_{\SC}(t) };
\end{flalign*}
with, 
\begin{flalign*}
    &\rho^{(\s 1)}_{\SC}(t) = \varrho^{(0,1)}_{\SC}(t), \\
    &\rho^{(1)}_{\SC}(t) = \varrho^{(0,1)}_{\SC}(t) + \varrho^{(1,0)}_{\SC}(t), \\
    &\rho^{(\s 2)}_{\SC}(t) = \varrho^{(0,2)}_{\SC}(t) + \varrho^{(1,1)}_{\SC}(t);
\end{flalign*}
where, as schematized in Fig. \ref{fig:flowchart}, the solvability condition solutions are intermediate orders in the multiple-scale perturbative expansion.
The three aforementioned trace distances oscillate with the frequency of the Rabi oscillations shown in Fig. \ref{fig:example_jc_populations:a}.
However, only the oscillations of $\mathcal{T}( \rho^{\num}_{\SC}, \rho^{(\s 2)}_{\SC}) $ are shown, whilst for $\mathcal{T}( \rho^{\num}_{\SC}, \rho^{(\s 1)}_{\SC}) $ and $\mathcal{T}( \rho^{\num}_{\SC}, \rho^{(1)}_{\SC})$ the curves follow the local maxima, since otherwise, the high frequency oscillations would make the data very difficult to visualize.
It is noteworthy that while the trace distance for short times decreases from the first solvability condition to the first-order solution; for long times, it increases, and needs the second-order solvability condition to decrease again. 
This behavior was evident in Figs. \ref{fig:example_jc_coherence}, where for short times the imaginary part of the coherence $\varrho^{(0,1)}_{1,2}(t)$ already corresponded with the exact dynamics; while for long times it differed by a constant factor, however, the first-order solution added imprecision that had to be corrected by the second-order solvability condition.
Besides, although not shown in Fig. \ref{fig:example_jc_td_sc}, the second-order solution reduces both the error for short and long times; though, the reduction is more significant for short times.
This irregular behavior of the error for short and long times to each order in the perturbative expansion, makes a quantitative estimation very difficult; nevertheless, as the trace distance describes very well what is observed for the quantum dynamics, we can conclude from a phenomenological analysis that trace distances below $0.001$ evidence excellent agreement of the approximation with the exact numerical results.
On the other hand, Fig. \ref{fig:example_jc_td_wc} shows the following trace distances for the WC regime,
\begin{flalign*}
    &\mathcal{T}\pqty{ \rho^{\num}_{\WC}(t), \rho^{(\s 2)}_{\WC}(t) },\,
    \mathcal{T}\pqty{ \rho^{\num}_{\WC}(t), \rho^{(2)}_{\WC}(t) };
\end{flalign*}
where, 
\begin{flalign*}
    &\rho^{(\s 2)}_{\WC}(t) = \varrho^{(0,2)}_{\WC}(t) + \varrho^{(1,1)}_{\WC}(t), \\
    &\rho^{(2)}_{\WC}(t) = \varrho^{(0,2)}_{\WC}(t) +     \varrho^{(1,1)}_{\WC}(t) + \varrho^{(2,0)}_{\WC}(t).     
\end{flalign*}
In this case, the trace distances for zero-order, first solvability condition, and first-order, are exactly the same and asymptotically approximately equal to $0.04$.
In this case the dynamics is much more regular and, as before, from a phenomenological description we can conclude that trace distances below $0.001$ correspond to very good perturbative approximations.
%
%-------------------------------------------------------------------------------
% Conclusions
\section{Conclusions}
\label{sec:conclusions}
To conclude, we have presented a multiple-scale perturbation technique that allows us to find excellent approximate solutions to time-local master equations describing open quantum systems. 
The technique provides the time-evolution of the corresponding dynamical map and, consequently, the time-evolution of the system density matrix for arbitrary initial conditions, allowing us to identify in each order the characteristic time scales involved in the problem.
The presented method is easy to implement using the general expressions derived in this paper, and is proper for a broad range of perturbations and open quantum systems.
Moreover, as is shown in the illustrative example, the method encourages separating the problem according to the physical regimes that are being described, facilitating the study of open quantum system from physical criteria.
Besides, if the solvable liouvillian and the perturbation commute, the presented technique reaches the exact solution in a single step in the perturbation theory.
The description in terms of the dynamical map is suitable to obtain analytical expressions for physical observables, such as the energy density spectrum, and high-order correlation functions; in a similar way, it is appropriate to calculate quantifiers and witnesses of entanglement, such as concurrence and negativity.
Finally, it is worth noting that the developed technique is a useful tool for the study of the dynamics of time-dependent quantum systems, both open and hamiltonian.
This, as consequence of the time-averaging procedure in the multiple-scale analysis, which allows us to deal with time-dependent parameters in a surprisingly simple way and, therefore, to find general analytical expressions for time-dependent quantum systems that would otherwise be very difficult to calculate.
In particular, the method is appropriate for systems from nonstationary cavity quantum electrodynamics, such as those related with the dynamical Casimir effect.
%
%-------------------------------------------------------------------------------

%------------------------------------------------------------------------------
% Acknowledgments

\section*{Acknowledgments}
    D.N.B.-G. and H.V.-P. acknowledge partial financial support from COLCIENCIAS under the projects,  ``Emisión en sistemas de qubits superconductores acoplados a la radiación'', code 110171249692, CT 293-2016, HERMES 31361; ``Control dinámico de la emisión en sistemas de qubits acoplados con cavidades no-estacionarias'', HERMES 41611; and ``Electrodinámica cuántica de cavidades no estacionarias'', HERMES 43351.
    D.N.B.-G. acknowledges funding from COLCIENCIAS through ``Convocatoria No. 727 de 2015 -- Doctorados Nacionales 2015''.
    B.A.R. acknowledges financial support from Universidad de Antioquia under Project No. 2014-989 (CODI-UdeA).
    B.A.R. and D.N.B.-G. gratefully acknowledge support from ``Proyecto de sostenibilidad del Grupo de Física Atómica y Molecular''.
    %
%
%------------------------------------------------------------------------------

%------------------------------------------------------------------------------
% Appendices
\appendix

%------------------------------------------------------------------------------
\section{Solution to the zero-order equation}
\label{app:homogeneous_equation}
It is worth noting, that given the nature of the mathematical objects involved in the zero-order equation, Eq. \eqref{eq:homogeneous_eqn}, the integration of the differential equation must be done in such a way that the temporal order is respected.  Therefore, we follow a procedure similar to the one used for the construction of the Dyson series.
From Eq. \eqref{eq:homogeneous_eqn}, we have,
\begin{align}
	\D_0\, \varrho^{(0)}(\T) - \L_0\, \varrho^{(0)}(\T) = 0.
\end{align}
Being $\varrho^{(0)}(\T) = \Phi^{(0)}(\T)\, \rho(0)$, we have then for the zero-order dynamical map,
\begin{align}
	\D_0\, \Phi^{(0)}(\T) - \L_0\, \Phi^{(0)}(\T) = \mathbb{0}.
\end{align}
Integrating with respect to $\tau_0$,
\begin{align}
    &\int_{0}^{\tau_0} \pdv{\Phi^{(0)}(\tau_0', \tau_1, \dots)}{\tau_0'} \dd{\tau_0'} = \nonumber \\
    & \hspace{4cm} \int_{0}^{\tau_0} \L_0\, \Phi^{(0)}(\tau_0', \tau_1, \dots)\,  \dd{\tau'_0} \nonumber \\
    &\Phi^{(0)}(\T)-\eval{\Phi^{(0)}(\tau'_0, \tau_1, \dots)}_{\tau'_0=0}  = \nonumber \\
    & \hspace{4cm} \L_0\, \int_{0}^{\tau_0} \Phi^{(0)}(\tau'_0, \tau_1, \dots)\,  \dd{\tau'_0}, \nonumber
\end{align}
where we label $\eval{\Phi^{(0)}(\tau'_0, \tau_1, \dots)}_{\tau'_0=0} = \C_{0,1}(\T_1)$. The indexing of the $\C_{0,1}(\T_1)$ superoperator is such that the first index is associated with the term of the expansion that is being determined, $\Phi^{(0)}$ in this case; and the second index is associated with the minimum-labeled time-scale on which the superoperator depends, here $\tau_1$. Hence, we have
\begin{equation}\label{eq:app_integral}
	\Phi^{(0)}(\T) = \C_{0,1}(\T_1) + \L_0 \int_{0}^{\tau_0} \Phi^{(0)}(\tau_0', \tau_1, \dots)\,  \dd{\tau'_0},
\end{equation}
then, we can solve this iteratively by replacing the integral form of $\Phi^{(0)}(\tau_0', \tau_1, \dots)$ in Eq. \eqref{eq:app_integral},
\begin{align}
	\Phi^{(0)}(\T) =& \C_{0,1}(\T_1) + \L_0\, \tau_0\, \C_{0,1}(T_1) \nonumber \\ &+ \L_0\, \int_{0}^{\tau_0} \int_{0}^{\tau_0'}  \Phi^{(0)}(\tau''_0, \tau_1, \dots)  \,  \dd{\tau''_0}  \,  \dd{\tau_0'},
\end{align}
with $\tau_0' > \tau_0''$. We continue in this way infinitely until the integration interval is so small that the integral is null and we get,
\begin{equation}
\Phi^{(0)}(\T) = \sum\limits_n \frac{1}{n!} \pqty{\L_0\, \tau_0}^n \C_{0,1}(\T_1)
\end{equation}
\begin{equation}
\Phi^{(0)}(\T) = \e^{\L_0 \tau_0} \C_{0,1}(\T_1),
\end{equation}
where there is an implicit time ordering associated with the fact that $\tau_0 > \tau_1 > ... > \tau_n $, which is guaranteed by the perturbative condition $0< \alpha < 1$.
Finally,
\begin{equation}
	\varrho^{(0)}(\T) = \Bqty{ \e^{\L_0 \tau_0} \C_{0,1}(\T_1) }\, \rho(0).
\end{equation}
%
%------------------------------------------------------------------------------

%------------------------------------------------------------------------------
\section{Commutation relations between the \texorpdfstring{$\K_n$}{e} superoperators}
\label{app:commutation}
Up to first-order in the MSPT, we have from Eq. \eqref{eq:f10} the definition of the superoperator $\F_{1,0}(\tau_0)$,
\begin{equation}
 \F_{1,0}(\tau_0) = \e^{-\K_0 \tau_0}\, \L_1(\tau_0)\, \e^{\K_0 \tau_0}, \quad \K_{1} = \expval{ \F_{1,0} }_{\tau_{0}}.
\end{equation}
It is clear that the equation of motion for $\F_{1,0}(\tau_0)$ is given by,
\begin{equation}
\D_0 \F_{1,0}(\tau_0) =  \comm{\F_{1,0}(\tau_0)}{\K_0}, \quad \F_{1,0}(0) = \L_1(0).
\end{equation}
Taking the time-average with respect to $\tau_0$, we have,
\begin{align}
\expval{ \D_0 \F_{1,0} }_{\tau_0} &= \comm{ \expval{ \F_{1,0} }_{\tau_0} }{\K_0} = \comm{\K_1\, }{\K_0\, },
\end{align}
\begin{align}
  \expval{ \D_0 \F_{1,0} }_{\tau_0}  &=\,  \frac{1}{T_{1,0}}\int_0^{T_{1,0}} \D_0 \F_{1,0}(\tau_0) \dd{\tau_0} \nonumber \\
  &=\, \frac{1}{T_{1,0}} \bqty{\F_{1,0}(T_{1,0})-\F_{1,0}(0)}{}\, =\, \mathbb{0},
\end{align}
since $\F_{1,0}$ is periodic with period $T_{1,0}$, and therefore,
\begin{equation}
    \comm{\K_1\, }{\K_0\, } = \mathbb{0}.
\end{equation}
Similarly, up to the $(n+1)$th-order in the MSPT ($n \geq 1$),
\begin{align}
    \F_{1,n}(\tau_n) =& \e^{-\K_n \tau_n} \expval{ \F_{2,n-1} }_{\tau_{n-1}} \e^{\K_n \tau_n}, \nonumber \\
    & \hspace{3.2cm} \K_{n+1} = \expval{ \F_{1,n} }_{\tau_{n}};
\end{align}
\begin{align}
    \D_n \F_{1,n}(\tau_n) =& \comm{\F_{1,n}(\tau_n)}{\K_n}, \quad \F_{1,n}(0) = \expval{\F_{2,n-1}}_{\tau_{n-1}}.
\end{align}
Analogously to the above, we take the time-average in a period $T_{1,n}$ of the superoperator $\F_{1,n}$,
\begin{align}
    \expval{ \D_n \F_{1,n} }_{\tau_n} &= \comm{ \expval{ \F_{1,n} }_{\tau_n} }{\K_n}, \quad \expval{ \D_n \F_{1,n} }_{\tau_n} = \mathbb{0};
\end{align}
and therefore,
\begin{equation}
    \comm{\K_{n+1}\, }{\K_n\, } = \mathbb{0}, \quad n \geq 0.
\end{equation}
This shows that the zeroth term in the asymptotic expansion of the dynamical map, $\Phi^{(0)}(\T) = \e^{\K_0 \tau_0}\,  \e^{\K_1 \tau_1}\, \dots \ \e^{\K_n \tau_n}\, \dots $, is an exponential factorization of commuting superoperators.
%
%------------------------------------------------------------------------------

%------------------------------------------------------------------------------
% Bibliography
%
%\section*{\refname} % commented for arXiv submit
%\bibliographystyle{elsarticle-harv}
\bibliographystyle{elsarticle-num} % 'Elsevier LaTeX' style
\bibliography{ms.bib}
%------------------------------------------------------------------------------

\end{document}